\documentclass{article}
\usepackage[utf8]{inputenc}
\usepackage{graphicx}
\usepackage{amssymb}
\usepackage{amsmath}
\newtheorem{theorem}{Theorem}[section]

\newtheorem{definition}{Definition}[section]

\newenvironment{proof}{{\noindent\it Proof}.\quad}{\hfill $\square$\par}
\usepackage{cmll}
\usepackage{booktabs}
\usepackage{dsfont}
\usepackage{caption}
\usepackage{cite}  % DO NOT CHANGE THIS AND DO NOT ADD ANY OPTIONS TO IT

\usepackage{dcolumn}% Align table columns on decimal point
\usepackage{bm}

\usepackage{mathptmx}
\usepackage{etoolbox}
\usepackage{booktabs}

\usepackage[utf8]{inputenc}
\usepackage[paperwidth=5.5in, paperheight=5.5in]{geometry}
\usepackage[T1]{fontenc}    % use 8-bit T1 fonts
\usepackage{hyperref}       % hyperlinks
\usepackage{url}            % simple URL typesetting
\usepackage{booktabs}       % professional-quality tables
\usepackage{amsfonts}       % blackboard math symbols
\usepackage{nicefrac}       % compact symbols for 1/2, etc.
\usepackage{microtype}      % microtypography
\usepackage{lipsum}		% Can be removed after putting your tex
\usepackage{doi}
\usepackage{indentfirst}
\usepackage{multirow}
\usepackage{float}
\usepackage{subfigure}
\usepackage{graphicx}
\usepackage{float}

\renewcommand{\thesubfigure}{(\roman{subfigure})}%此外，还可设置图编号显示格式，加括号或者不加括号
\makeatletter \renewcommand{\@thesubfigure}{\thesubfigure \space}%子图编号与名称的间隔设置

\begin{document}

\title{Efficient Privacy-Preserving Convolutional Spiking Neural Networks with FHE}
\maketitle

\author{Pengbo Li, Huifang Huang, Ting Gao\thanks{*}, Jin Guo, Jinqiao Duan}

\begin{abstract}
With the rapid development of AI technology, we have witnessed numerous innovations and conveniences. However, along with these advancements come privacy threats and risks. Fully Homomorphic Encryption (FHE) emerges as a key technology for privacy-preserving computation, enabling computations while maintaining data privacy. Nevertheless, FHE has limitations in processing continuous non-polynomial functions as it is restricted to discrete integers and supports only addition and multiplication operations. Spiking Neural Networks (SNNs) operate on discrete spike signals, naturally aligning with the properties of FHE.

In this paper, we present a framework called FHE-DiCSNN. This framework is based on the efficient TFHE scheme and leverages the discrete properties of SNNs to achieve remarkable prediction performance on ciphertext (up to 97.94\% accuracy) with a time efficiency of 0.75 seconds per prediction. Firstly, by employing bootstrapping techniques, we successfully implement computations of the Leaky Integrate-and-Fire (LIF) neuron model on ciphertexts. Through bootstrapping, we can facilitate computations for SNNs of arbitrary depth. This framework can be extended to other spiking neuron models,  providing a novel framework for the homomorphic evaluation of SNNs. Secondly, inspired by Convolutional Neural Networks (CNNs), we adopt convolutional methods to replace Poisson encoding. This not only enhances accuracy but also mitigates the issue of prolonged simulation time caused by random encoding. Furthermore, we employ engineering techniques to parallelize the computation of bootstrapping, resulting in a significant improvement in computational efficiency. Finally, we evaluate our model on the MNIST dataset. Experimental results demonstrate that, with the optimal parameter configuration, FHE-DiCSNN achieves an accuracy of 97.94\% on ciphertexts, with a loss of only 0.53\% compared to the original network's accuracy of 98.47\%. Moreover, each prediction requires only 0.75 seconds of computation time.
\end{abstract}

\section{Introduction}

\textbf{Privacy-Preserved AI.}
In recent years, privacy preservation has garnered significant attention in the field of machine learning. Fully Homomorphic Encryption (FHE) has emerged as the most suitable tool for facilitating Privacy-Preserving Machine Learning (PPML) due to its robust encryption security and efficient communication capabilities. The foundation of FHE was established in 2009 when Gentry introduced the first fully homomorphic encryption scheme \cite{gentry2009fully, gentry2010computing} capable of evaluating arbitrary circuits. His pioneering work not only proposed the FHE scheme but also outlined a method for constructing a comprehensive FHE scheme from a model with limited yet sufficient homomorphic evaluation capacity. Inspired by Gentry's groundbreaking contributions, subsequent second-generation schemes like BGV \cite{brakerski2014leveled} and FV \cite{fan2012somewhat} had been proposed. The evolution of FHE schemes had continued with third-generation models such as FHEW \cite{ducas2015fhew}, TFHE \cite{chillotti2020tfhe}, and Gao \cite{case2019fully, gao2018efficient}, which had offered rapid bootstrapping and had supported an unlimited number of operations. The CKKS scheme \cite{cheon2017homomorphic, cheon2019full} had attracted considerable interest as a suitable tool for PPML implementation, given its natural handling of encrypted real numbers.

However, existing FHE schemes had primarily supported arithmetic operations such as addition and multiplication, while widely used activation functions such as ReLU, sigmoid, leaky ReLU, and ELU had been non-arithmetic functions. To overcome this challenge, Dowlin et al. \cite{gilad2016cryptonets} introduced CryptoNets, which utilized neural networks, particularly artificial feedforward neural networks trained on plaintext data, to provide accurate predictions on homomorphically encrypted data. Nonetheless, CryptoNets had faced performance limitations to some extent due to the replacement of the sigmoid activation function and associated computational overhead. Zhang et al. \cite{zhang2017privacy} had proposed a privacy-preserving deep learning model called the dual-projection deep computation model, which utilized cloud outsourcing to enhance learning efficiency and combined it with the BGV scheme for training. Building upon CryptoNets, Brutzkus et al. \cite{brutzkus2019low} had developed an enhanced version that had reduced latency and optimized memory usage. Furthermore, Lee et al. \cite{lee2022privacy} demonstrated the potential of applying FHE (with bootstrapping) to deep neural network models by implementing ResNet-20 on the CKKS scheme.

In a distinct study \cite{bourse2018fast}, the authors developed the FHE-DiNN framework, a discrete neural network framework predicated on the TFHE scheme. Unlike traditional neural networks, FHE-DiNN had discretized network weights into integers and utilized the sign function as the activation function. The computation of the sign function had been achieved through bootstrapping on ciphertexts. Each neuron's output had been refreshed with noise, thereby enabling the neural network to extend computations to any depth. Although FHE-DiNN offered high computational speed, it had compromised model prediction accuracy. Given the close resemblance between the sign function and the output of Spiking Neural Network(SNN) neurons, this work provided a compelling basis for investigating efficient homomorphic evaluations of SNNs in the context of PPML.

\textbf{CNNs and SNNs.} Convolutional Neural Networks (CNNs) have emerged as powerful tools in the field of computer vision, offering exceptional accuracy and an automated feature extraction process \cite{dhillon2020convolutional}. The unique structures of CNNs, including convolutional and pooling layers, are built upon three key concepts: (a) local receptive fields, (b) weight sharing, and (c) spatial subsampling. These elements eliminate the need for explicit feature extraction (Convolutional layer) and reduce training time, making CNNs highly suitable for visual recognition tasks \cite{voulodimos2018deep}. In recent years, CNNs have found widespread applications in various domains, such as image classification and recognition \cite{simonyan2014very,szegedy2015going,zeiler2014visualizing}, Natural Language Processing (NLP) \cite{bradbury2016quasi,lee2017fully}, object detection \cite{kim2016pvanet}, and video classification \cite{ballas2015delving,mathieu2015deep}. The widespread adoption of CNNs has played a significant role in the advancement of deep learning.

Spiking Neural Networks (SNNs), regarded as the third generation of neural networks \cite{Maass1996NetworksOS}, operate in a manner more akin to biological reality compared to their predecessors. Unlike the widespread use of Artificial Neural Networks (ANNs), SNNs uniquely process information in both space and time, capturing the temporal dynamics of biological neurons. Neuron models, the fundamental units of SNNs, have been constructed by neurophysiologists in numerous forms. Among these, the most influential models include the Hodgkin-Huxley (H-H) model \cite{Hodgkin1952AQD}, the leaky integrate-and-fire (LIF) model \cite{Wu2017SpatioTemporalBF}, the Izhikevich model \cite{Izhikevich2003SimpleMO}, and the spike response model \cite{jolivetSpikeResponseModel2003} (SRM). These models are distinguished by their use of spikes or 'action potentials' for information communication, closely emulating the behavior of neurons in the brain. This temporal aspect of information processing enables SNNs to manage time-series data more naturally and efficiently than traditional artificial neural networks.

Convolution Spiking Neuron Networks (CSNNs) represent the integration of these two powerful models into CSNNs which brings together the spatial feature learning capabilities of CNNs and the temporal dynamic processing of SNNs. This combination allows CSNNs to process spatiotemporal data more efficiently and accurately, making them particularly suitable for tasks such as video processing, speech recognition, and other real-time sensory data processing tasks. Zhou et al. \cite{zhou2021temporal} built upon \cite{mostafa2017supervised} to create a sophisticated architecture for SNNs, utilizing the VGG16 model for CIFAR10 \cite{liu2015very,simonyan2014very} and the GoogleNet model for ImageNet \cite{szegedy2015going}. In parallel, Zhang et al. \cite{zhang2021rectified} devised a deep convolutional spiking neural network encompassing two convolutional layers and two hidden layers, employing a ReL-PSP-based spiking neuron model and training the network through temporal BP with recursive backward gradients. Further, a range of CSNNs works \cite{kundu2021spike,muramatsu2023combining,sengupta2019going,rueckauer2018conversion} represent converted variants of conventional CNNs, while others \cite{zhang2020temporal,lee2020enabling} incorporate BP directly onto the network using rate coding or multi-peak per neuron strategies.

Traditional neural networks rely on real numbers for computations, with neurons' outputs and network weights represented as continuous values. However, homomorphic encryption algorithms are unable to directly operate on real numbers. Consequently, in order to perform homomorphic computations, the outputs, and weights of the neural network must be discretized into integers. In contrast, Discrete Convolutional Spiking Neural Networks (DiCSNNs) are characterized by neuron outputs that fundamentally consist of discrete value signals, necessitating only the discretization of weights. From this standpoint, SNNs demonstrate greater suitability for homomorphic computations compared to traditional neural networks.

\textbf{Our Contribution.} 
In this paper, we propose the FHE-DiCSNN framework. Built upon the efficient TFHE scheme and incorporating convolutional operations from CNN, this framework harnesses the discrete nature of SNNs to achieve exceptional prediction performance on ciphertexts (with a maximum accuracy of 97.94\%) while maintaining a time efficiency of 0.75 seconds per prediction.

1. By successfully implementing the FHE-Fire and FHE-Reset functions using bootstrapping techniques, we enable computations of LIF neurons on ciphertexts. This approach can be extended to other SNNs neuron models, offering a novel solution for privacy protection in third-generation neural networks (SNNs).

2. LIF neurons serve as activation functions in deep networks, forming the Spiking Activation Layer. The bootstrapped LIF model generates ciphertext with minimal initial noise. By ensuring that the accumulated noise after linear layer operations remains below a predefined threshold, subsequent layers in the Spiking Activation Layer share the same initial noise, enabling further computations. Our framework allows the network to expand to any depth without noise-related concerns.

3. To convert signals into spikes, we replace Poisson encoding with convolutional methods. This not only enhances accuracy but also mitigates the issue of prolonged simulation time caused by randomness. Additionally, we employ engineering techniques to parallelize the bootstrapping computation, resulting in a significant improvement in computational efficiency.

We conducted experiments on the MNIST dataset to validate the advantages of FHE-DiCSNN. Firstly, using the Spikingjelly package, we trained CSNNs with different parameters, including LIF and IF models. The results indicate that the decay factor $\tau$ of the LIF model significantly affects accuracy. Next, we discretized the trained network and determined FHE parameters based on experimental results and theoretical analysis. Finally, we evaluated the accuracy and time efficiency of the FHE-DiCSNN framework on ciphertexts. Experimental results demonstrate that, with optimal parameter configuration, FHE-DiCSNN achieves a ciphertext accuracy of 97.94\%, with only a 0.53\% loss compared to the original network's accuracy of 98.47\%. Moreover, each prediction requires only 0.75 seconds of computation time.

\textbf{Outline of the paper.}
The paper is structured as follows: Section 2 provides definitions and explanations of SNNs and TFHE, including a brief introduction to the bootstrapping process of TFHE. In Section 3, we present our method of constructing Discretized Convolutional Spiking Neural Networks and prove that the discretization error can be controlled. In Section 4, we highlight the challenges of evaluating DiCSNN homomorphically and provide a detailed explanation of our proposed solution. In Section 5, we present comprehensive experimental results for verification of our proposed framework. And discuss the challenges and possible future work in section 6.

\section{Preliminary Knowledge}

In this section, we aim to offer a comprehensive elucidation of the bootstrapping operations within the framework of the TFHE scheme. Additionally, we will provide an in-depth exposition of the background knowledge pertinent to Spiking Neural Networks (SNNs).

\subsection{Programmable Bootstrapping}
\newtheorem{theorem1}{Theorem}[subsection]
\newtheorem{remark1}[theorem1]{Remark}
\newtheorem{proposition1}[theorem1]{Proposition}
\newtheorem{corollary1}[theorem1]{Corollary}
\newtheorem{definition1}[theorem1]{Definition}

Firstly, let us define some mathematical symbols and concepts used in FHE.

Set $\mathbb{Z}_p = \left\{-\frac{p}{2}+1, \ldots, \frac{p}{2}\right\}$ denote a finite ring defined over the set of integers. The message space for homomorphic encryption is defined within this finite ring $\mathbb{Z}_p$. 

Consider $N=2^k$ and the cyclotomic polynomial $X^N+1$, then
$$
R_{q, N} \triangleq R / q R \equiv \mathbb{Z}_q[X] /\left(X^N+1\right) \equiv \mathbb{Z}[X] /\left(X^N+1, q\right).
$$
Similarly, we can define the polynomial ring $R_{p, N}$.

Before discussing the programmable bootstrapping theorem, we will introduce three homomorphic encryption schemes used.
\begin{itemize}
\item \noindent\textbf{LWE (Learning With Errors).}We revisit the encryption form of LWE \cite{regev2009lattices} as shown in Figure \ref{ZP}, which is employed to encrypt a message $m \in \mathbb{Z}_p$ as
$$LWE_s(m)=(\mathbf{a}, b)=\left(\mathbf{a},\langle\mathbf{a}, \mathbf{s}\rangle+e+ \left\lfloor\frac{q}{p} m\right\rceil\right) \bmod q, $$
where $\mathbf{a} \in \mathbb{Z}_q^n$, $b \in \mathbb{Z}_q$, and the keys are vectors $\mathbf{s} \in \mathbb{Z}_q^n$. The ciphertext $(\mathbf{a}, b)$ is decrypted using:
$$
\left\lfloor \frac{p}{q}(b-\langle\mathbf{a}, \mathbf{s}\rangle ) \right\rceil \bmod p=\left\lfloor m+\frac{p}{q} e\right\rceil=m
.$$

\begin{figure}[htb]
    \centering
     \label{ZP}
    \includegraphics[width=0.4\textwidth]{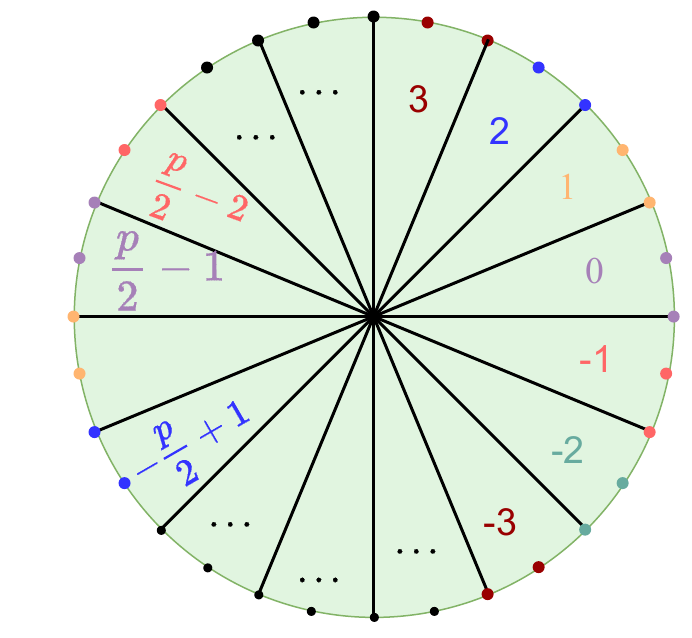}
    \caption{The partitioning of the circle serves to reflect the mapping relationship between $\mathbb{Z}_p$ and $\mathbb{Z}_q$.}
    \label{ZP}
\end{figure}

\item \noindent\textbf{RLWE (Ring Learning With Errors)\cite{lyubashevsky2010ideal}.} An RLWE ciphertext of a message $m(X) \in R_{p,N}$ can be obtained as follows:
$$R L W E_s(m(X)) =  \left(a(X), b(X)\right), \text{where } b(X) = a(X) \cdot s(X)+e(X)+ \left\lfloor \frac{q}{p} m(X) \right\rceil,$$
where $a(X) \leftarrow R_{q,N}$ is uniformly chosen at random, and $e(X) \leftarrow \chi_\sigma^n$ is selected from a discrete Gaussian distribution with parameter $\sigma$. The decryption algorithm for RLWE is similar to LWE.

\item \noindent\textbf{GSW.}
As one of the third-generation fully homomorphic encryption schemes, the GSW\cite{gentry2013homomorphic} scheme exhibits advantages in both efficiency and security. Furthermore, its variant, RGSW\cite{gentry2013homomorphic}, has been widely applied in practical scenarios. Given a plaintext $m \in \mathbb{Z}_p$, the plaintext $m$ is embedded into a power of a polynomial to obtain $X^m \in R_{p,N}$, which is then encrypted as $RGSW(X^m)$. RGSW enables efficient computation of homomorphic multiplication, denoted as $\diamond$, while effectively controlling noise growth:
$$
\begin{aligned}
& R G S W\left(X^{m_0}\right) \diamond R G S W\left(X^{m_1}\right)=R G S W\left(X^{m_0+m_1}\right),\\
& R L W E\left(X^{m_0}\right) \diamond R G S W\left(X^{m_1}\right)=R L W E\left(X^{m_0+m_1}\right).
\end{aligned}
$$
\end{itemize}

Now, we present the theorem of Programmable Bootstrapping.
\begin{theorem}\label{Programtfhe}
(Programmable Bootstrapping\cite{chillotti2021improved}) TFHE/FHEW bootstrapping enables the computation of any function $g$ with the property $g: \mathbb{Z}_p \rightarrow \mathbb{Z}_p$ and $g\left(v+\frac{p}{2}\right)=-g(v)$. The function $g$ is referred to as the program function of bootstrapping. Given an LWE ciphertext $LWE(m)_s=(\mathbf{a}, b)$, where $m \in \mathbb{Z}_p$, $\mathbf{a} \in \mathbb{Z}_p^N$, and $b \in \mathbb{Z}_p$, it is possible to bootstrap it into $L W E_s(g(m))$ with very low initialization noise.
\end{theorem}

The bootstrapping process relies on the Homomorphic Accumulator denoted as $A C C_g$. Utilizing the notations from\cite{micciancio2021bootstrapping}, the bootstrapping process can be divided into the following steps:
\begin{itemize}
\item\textbf{Initialization.} Obtain the initial polynomial by:
\begin{equation*}
    A C C_g[-b]=X^{-b} \cdot \sum_{i=0}^{N-1} g\left(\left\lfloor\frac{i \cdot p}{2 N}\right\rfloor\right) X^i \bmod X^N+1 .
\label{Initialization}
\end{equation*}

\item\textbf{Blind Rotation.} The $ACC_g \leftarrow_{+}^{+}-a_i \cdot e k_i$ modifies the content of the accumulator from $ACC_g[-b]$ to $ACC_g\left[-b+\sum a_i s_i\right]=ACC_g[-m-e]$, where
\begin{equation*}
    \text { ek }=\left(R G S W\left(X^{s_1}\right), \ldots, R G S W\left(X^{s_n}\right)\right), \text { over } R_q^N.
    \label{Blind_Rotation}
\end{equation*}

\item\textbf{Sample Extraction.} The $ACC_g=(a(X), b(X))$ is the RLWE ciphertext with component polynomials $a(X)=\sum_{0 \leq i \leq N-1} a_i X^i$ and $b(X)=\sum_{0 \leq i \leq N-1} b_i X^i$. The extraction operation outputs the LWE ciphertext:
% \begin{equation}
%     R L W E_z \stackrel{\text { Sample Extraction }}{\longrightarrow} L W E_z(g(m))=\left(\mathbf{a}, b_0\right),
%     \label{Sample_Extraction}
% \end{equation}
\begin{equation*}
    R L W E_z \xrightarrow{\text{ Sample Extraction}} L W E_z(g(m))=\left(\mathbf{a}, b_0\right),
    \label{Sample_Extraction}
\end{equation*}
where $\mathbf{a}=\left(a_0, \ldots, a_{N-1}\right)$ is the coefficient vector of $a(X)$, and $b_0$ is the coefficient of $b(X)$.

\item\textbf{Key Switching.} Key switching transforms the key of the LWE instance from the original vector $\mathbf{z}$ to the vector $\mathbf{s}$ while preserving the plaintext message $m$:
\begin{equation*}
    L W E_{\mathbf{z}}(g(m)) \xrightarrow{\text { Key Switching }} L W E_{\mathbf{s}}(g(m)).
    \label{Key_Switching}
\end{equation*}

\end{itemize}

By utilizing a bootstrapping key and a KeySwitching key as input, bootstrapping can be defined as follows:
$$
\text { bootstrapping }=\text { KeySwitch } \circ \text { Extract } \circ \text { BlindRotate } \circ \text { Initialize. }
$$

Given a program function $g$, bootstrapping is a process that takes an LWE ciphertext $LWE_s(m)$ as input and outputs $LWE_s(g(m))$ with the original secret key $s$:
\begin{equation*}
    \text { bootstrapping }\left(L W E_s(m)\right)=L W E_s(g(m)) \text {. }
\end{equation*}

This property will be extensively utilized in our context. Since bootstrapping does not modify the secret key, we will use the shorthand $LWE(m)$ to refer to an LWE ciphertext throughout the rest of the text.

\subsection{ Leaky Integrate-and-Fire Neuron Model}

Neurophysiologists have developed a range of models to capture the dynamic characteristics of neuronal membrane potentials, which are essential for constructing SNNs and determining their fundamental dynamical properties. Prominent models that have had a significant impact on neural networks include the Hodgkin-Huxley (H-H) model \cite{hodgkin1952quantitative}, the leaky integrate-and-fire (LIF) model \cite{wu2018spatio}, the Izhikevich model \cite{izhikevich2003simple}, and the spike response model \cite{jolivet2003spike} (SRM), among others. In this study, we selected the leaky integrate-and-fire (LIF) model, as shown in Eq.(\ref{LIF_model}), as the primary focus. This choice was made due to the simplicity of the LIF model and its ability to effectively describe the dynamic behavior of biological neurons.

\begin{equation}
    \tau \frac{\mathrm{d} V}{\mathrm{~d} t}=V_{\text{rest}}-V+RI,
    \label{LIF_model}
\end{equation}
where $\tau$ represents the membrane time constant indicating the decay rate of the membrane potentials, $V_{\text{rest}}$ represents the resting potential, the $R$ and $I$ terms denote the membrane impedance and input current, respectively.

The LIF model greatly simplifies the process of action potentials while retaining three key features of actual neuronal membrane potentials: leakage, accumulation, and threshold excitation. Building upon this foundation, there exists a series of variant models, such as the second-order LIF model\cite{brunel2003firing}, exponential LIF model\cite{fourcaud2003spike}, adaptive exponential LIF model\cite{brette2005adaptive}, and others. These variant models focus on describing the details of neuronal pulse activity and further enhance the biological plausibility of the LIF model at the cost of additional implementation complexity.

In practical applications, it is common to employ discrete difference equations as an approximation method for modeling the equations governing neuronal electrical activity. While the specific accumulation equations for various neuronal membrane potentials may vary, the threshold excitation and reset equations for the membrane potential remain consistent. Consequently, the neuronal electrical activity can be simplified into three distinct stages: charging, firing, and resetting, depicted as follows:
\begin{equation}
\label{LIF_equation}
\left\{
\begin{aligned}
    &H[t] = V[t-1] + \frac{1}{\tau}(-(V[t-1] - V_{\text{reset}}) + I[t]),\\
    &S[t] = \text{Fire}\left(H[t]-V_{\text{th}}\right),\\
    &V[t] = \operatorname{Reset}(H[t]) = \begin{cases}
       V_{\text{reset}}, & \text{if } H[t] \geq V_{\text{th}}; \\
       H[t], & \text{if } V_{\text{reset}} \leq H[t] \leq V_{\text{th}}; \\
       V_{\text{reset}}, & \text{if } H[t] \leq V_{\text{reset}}.
    \end{cases}
\end{aligned}\right.
\end{equation}

Generally set $R = 1$ and the $\text{Fire}(\cdot)$ is a step function defined as:
\begin{equation*}
    \operatorname{Fire}(x)= \begin{cases}1, & \text { if } \quad x \geq 0; \\ 0, & \text { if } \quad x \leq 0. \end{cases}
\end{equation*}

% 在这里，$ I[t] = \sum_j \omega_{j} x_j[t]$ 表示从突触前神经元或图像像素传入的外部输入的总膜电流。每个LIF神经元的输入通过加权求和计算得到。这种计算可以在卷积层或线性层（全连接层）中进行，因为这两种操作都涉及计算加权输入的总和，即WeightSum。
% 本文中，$x_j$表示相应的输入值，$\omega_j$表示与每个输入相关的权重。在$\sum_j \omega_j$符号中，j表示该层神经元的数量，当$\omega_j$表示全连接网络的参数时；j表示对应滤波器尺寸的平方，当$\omega_j$表示卷积层或平均池化层的参数时。我们将在后文中继续使用这些符号。

Here, the equation $ I[t] = \sum_j \omega_{j} x_j[t]$ denotes the cumulative membrane current resulting from external inputs originating from pre-synaptic neurons or image pixels. The input of each Leaky Integrate-and-Fire (LIF) neuron is obtained through a weighted sum calculation. This computational process can be performed in either convolutional layers or linear layers (fully connected layers), as both operations involve the calculation of the weighted sum of inputs, referred to as WeightSum.

In this context, $x_j$ represents the respective input value, while $\omega_j$ corresponds to the weight associated with each input. Within the notation $\sum_j \omega_j$, the variable j represents the number of neurons in the layer when $\omega_j$ represents the parameters of a fully connected network. Conversely, when considering convolutional layers or average pooling layers, j represents the square of the corresponding filter size when $\omega_j$ denotes the parameters. These symbols will continue to be utilized in subsequent discussions.

\subsection{Spiking Neural Networks}
Due to the non-differentiable nature of spikes \cite{zhang2018plasticity}, the conventional backpropagation (BP) algorithm cannot be directly applied to SNNs \cite{pfeiffer2018deep}. Training SNNs is a captivating research direction, there are some commonly used training methods such as ANN-to-SNN conversion and unsupervised training with STDP, and the gradient surrogate method is adopted for training SNNs in this study.

The main idea is to use a similar continuous function to replace the spike function or its derivative, resulting in a spike-based BP algorithm. Wu et al. \cite{Wu2017SpatioTemporalBF} introduce four curves to 
approximate the derivative of spike activity denoted by $f_1, f_2, f_3, f_4$ as follow:
$$
\begin{aligned}
& f_1(V)=\frac{1}{a_1} \operatorname{sign}\left(\left|V-V_{th}\right|<\frac{a_1}{2}\right), \\
& f_2(V)=\left(\frac{\sqrt{a_2}}{2}-\frac{a_2}{4}\left|V-V_{t h}\right|\right) \operatorname{sign}\left(\frac{2}{\sqrt{a_2}}-\left|V-V_{t h}\right|\right), \\
& f_3(V)=\frac{1}{a_3} \frac{e^{\frac{V_{th}-V}{a_3}}}{\left(1+e^{\frac{V_{th}-V}{a_3}}\right)^2}, \\
& f_4(V)=\frac{1}{\sqrt{2 \pi a_4}} e^{-\frac{\left(V-V_{t h}\right)^2}{2 a_4}}.
\end{aligned}
$$

In general, the training of SNNs adheres to three fundamental principles: (1) Spiking neurons generate binary output that is susceptible to noise. The temporal firing frequency serves as a representative measure of the strength of category responses for classification tasks. (2) The primary objective is to ensure that only the correct neuron fires at the highest frequency, while other neurons remain quiescent. Mean Squared Error (MSE) loss is frequently employed for training, as it has demonstrated enhanced performance. (3) Resetting the network state after each simulation is crucial.

Moreover, SNNs exhibit suboptimal performance in handling real-world data, such as image pixels and floating-point values. To address various stimulus patterns effectively, SNNs commonly employ a range of encoding methods, including rate coding, temporal coding, bursting coding, and population coding\cite{georgopoulos1986neuronal}, to process input stimuli. In our study, the inputs are encoded into rate-based spike trains by the Poisson process, name Poisson encoding. Given a time interval $\Delta t$ in advance, then the reaction time is divided into $T$ intervals evenly. During each time step $t$, a random matrix $M_t$ is generated using uniform distribution in $[0,255]$. Then, we compare the original normalized pixel matrix $X_o$ with $M_t$ to determine whether the current time $t$ has a spike or not. The final encoding spike $X$ is calculated by using the following equation:
\begin{equation*}\label{Poisson-coding}
   X(i,j)= 
   \begin{cases}
   0, &\ X_o(i,j) \leq M_t(i,j),\\
   1, &\ X_o(i,j) > M_t(i,j),
   \end{cases}
\end{equation*}
where $i$ and $j$ are the coordinates of the pixel points in the images. In this way, the encoded spikes follow the Poisson distribution.

\section{Discretized Convolutional Spiking Neural Network}
\subsection{Convolutional Spiking Neural Networks}
Convolutional Neural Networks (CNNs) capitalize on the local perception and weight-sharing characteristics inherent in convolution operations, enabling efficient extraction of image features using a limited number of convolution kernels. Consequently, CNN is capable of more effectively extracting and learning features from images without relying on the complexity and high computational costs associated with random coding.

In contrast, Poisson encoding serves as a simple random coding technique employed to convert continuous signals into pulse signals. However, Poisson encoding itself lacks the capability to extract image features. Due to its stochastic nature, employing Poisson encoding for signal encoding necessitates a large number of pulse samples to ensure the preservation of relevant information. This, in turn, leads to longer simulation times required for accurate extraction of image features, thereby increasing computational costs and time overhead.

% CNNs consist of essential layers: input, convolutional, activation function, pooling, and fully connected layers. The convolutional layer extracts valuable features, the activation function layer enables non-linear modeling, the pooling layer maintains scale invariance and reduces overfitting, and the fully connected layer integrates local features into global representations.

% Traditional neural networks with continuous activation functions like Tanh and Sigmoid fail to accurately simulate the dynamics of biological neurons. Sparse Neural Networks (CSNNs) address this issue by modifying the neuron model and learning mechanisms to achieve sparsity. The Leaky Integrate-and-Fire (LIF) neuron model (including IF and QIF) serves as a biologically interpretable activation function, converting input currents into binary spikes. Spike information, characterized by timing and frequency, provides valuable insights. Neurons require reaching a threshold for information propagation, but additional information can be inferred from spike timing and frequency.

\begin{figure}[htbp]
    \centering
		\centering
		% \vspace{-0.3cm}
		% \setlength{\abovecaptionskip}{0.28cm}
		\includegraphics[width=1\textwidth]{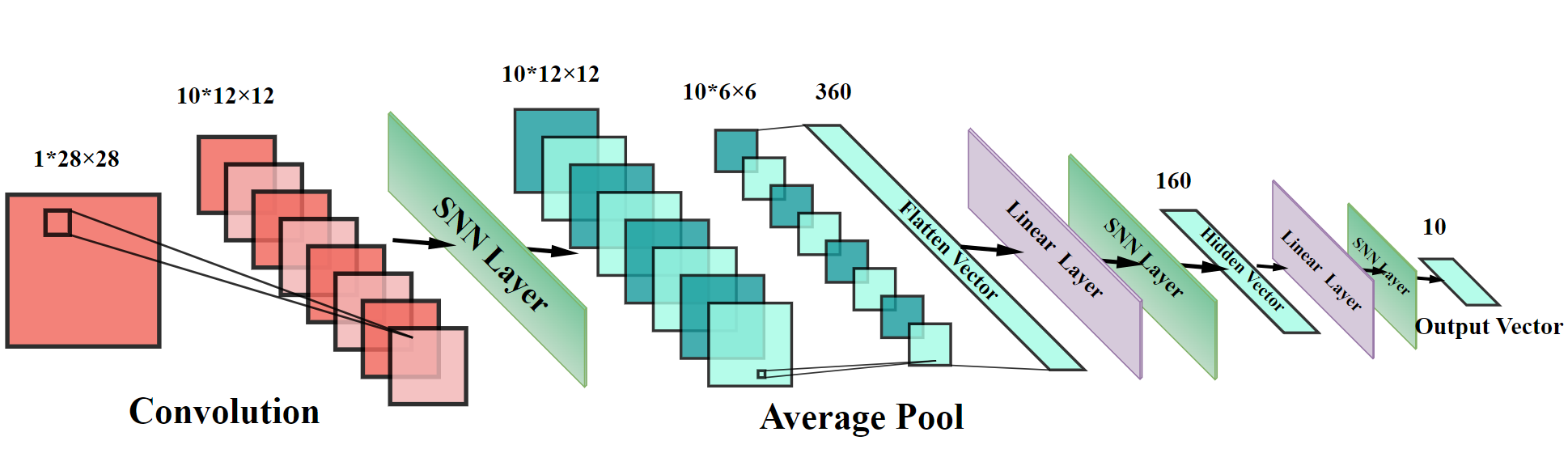}
        \captionsetup{font=small}
        \caption{The visualization diagram of the CSNNs network. In the proposed model, different colors are employed to signify distinct layers: red designates the convolutional layer, cyan is utilized for the pooling layer, green also denotes a separate convolutional layer, and purple illustrates the linear layer.}
    \label{CSNN_main}
\end{figure}

CSNNs (Convolutional Spiking Neural Networks) is a neural network model that combines CNNs and SNNs. In CSNNs, the LIF model or other spiking models, as mentioned earlier, are used to simulate the electrical activity of neurons, forming the Spiking Activation Layer in the network. Combined with convolutional layers, CSNNs extract image features and encode them into spike signals. By leveraging the spatial feature extraction capability of CNN and the spiking transmission characteristics of SNNs, CSNNs benefit from both convolution operations and discrete spike transmission. The visualization diagram of a CSNNs is presented in Figure \ref{CSNN_main}, and described in detail as follows:\\
\begin{itemize}
    \item Convolutional layer: The input image has a size of $28\times28$ with a padding dimension of 1. The convolution window or kernel size is $8\times8$ with a stride of $(2,2)$, resulting in 10 feature maps. Consequently, the output size of this layer is $10\times12\times12$.
   \item Spiking Activation Layer: Each input node is activated using the LIF neuron model.
   \item Scale average pooling layer: This layer applies a window size of $2\times2$, leading to an output size of $10\times6\times6$.
   \item Fully connected layer(Linear layer): This layer connects the 360 input nodes to the 160 output nodes, which is equivalent to performing matrix multiplication with a $160\times360$ matrix.
   \item Spiking Activation Layer: Each input node is activated using the LIF neuron model.
   \item Fully connected layer(Linear layer): This layer connects the 160 input nodes to the 10 output nodes.
   \item Spiking Activation Layer: The LIF neuron activation is applied to each of the 10 input values.
\end{itemize}

\subsection{Discretized CSNN}
%%%%%%%%%%%%%%%%%%%%%%%%%%%%%%%%%%%%%%%%一会就来
Traditional neural networks rely on real numbers for computations, with neurons' outputs and network weights represented as continuous values. However, homomorphic encryption algorithms are unable to directly operate on real numbers. Consequently, in order to perform homomorphic computations, the outputs, and weights of the neural network must be discretized into integers. In contrast, discretized CSNNs are characterized by neuron outputs that fundamentally consist of discrete value signals, necessitating only the discretization of weights. From this standpoint, CSNNs demonstrate greater suitability for homomorphic computations compared to traditional neural networks.

% Traditional neural networks rely on continuous computations with real numbers, whereas spiking neurons utilize discrete spike signals. As a result, CSNN benefits from the use of discrete weight values, making it particularly well-suited for homomorphic evaluation.
%%%%%%承上启下介绍DEF1
\begin{definition}
A Discretized Convolutional Spiking Neural Network (DiCSNNs) is characterized as a feed-forward spiking neural network wherein all weights, inputs, and outputs of the neuron model undergo discretization, resulting in their representation as elements of a finite set $\mathbb{Z}_p$, which signifies the integers modulo $p$.
\end{definition}
  
We utilize fixed-precision real numbers and apply suitable scaling to convert the weights into integers, effectively discretizing CSNNs into DiCSNNs. Denote this discretization method as the following function:

$$
\hat{x} \triangleq \operatorname{Discret}(x, \theta)=\lfloor x \cdot \theta\rceil,
$$
where $\theta \in \mathbb{Z}$ is referred to as the scaling factor, and $\lfloor \cdot \rceil$ represents rounding to the nearest integer. The discretized result of $x$ is denoted as $\hat{x}$. Moreover, alternative methodologies exist to accomplish this objective. Within the encryption process, all relevant numerical values are defined on the finite ring $\mathbb{Z}_p$. Hence, it is imperative to carefully monitor the numerical fluctuations throughout the computation to prevent reductions modulo $p$, as such reductions could give rise to unanticipated errors in the computational outcomes.

% Before the discretization of CSNNs, we note that 
% the division is an operation that should be avoided for FHE being poor with it. And it is more suitable to use the Sign function which outputs $\{-1, 1\}$ for implementing the Fire function for programmable bootstrapping. As a result, it is necessary to transform the equation into an equivalent form to accommodate this requirement.

It is important to highlight that the computation of LIF neurons is influenced by the discretization of weights, necessitating appropriate modifications. The essence of discretization lies in accommodating the requirements of FHE. Specifically, we address two aspects in this regard.

Firstly, the initial equation of LIF neurons (Equation \ref{LIF_equation}) involves a division operation, which poses challenges in the context of computation. Therefore, it is imperative to find alternative approaches that avoid explicit division calculations, ensuring compatibility with FHE.

Secondly, the Fire function in LIF neurons, representing a step function, relies on the programming function $g$ employed in bootstrapping techniques. To satisfy the condition $g(x) = -g(\frac{p}{2} + x)$, the Sign function proves to be more suitable than the Fire function. Therefore, it is worth considering using the Sign funnction as a replacement for the Fire function.

To provide comprehensive insights, we present Theorem \ref{discret_LIF}, which elucidates the discretization process of LIF neuron weights.

\begin{theorem}\label{discret_LIF}
Under the given conditions $V_{reset} = 0$ and $V_{th} = 1$, Eq.(\ref{LIF_equation}) can be discretized into the following equivalent form with a discretization scaling factor $\theta$:

\begin{equation}
\left\{
\begin{aligned}
& \hat{H}[t]=\hat{V}[t-1]+ \hat{I}[t], \\
& 2S[t]=\text { Sign }\left(\hat{H}[t]-\hat{V}_{\text {th}}^{\tau}\right) +1, \\
& \hat{V}[t]=\operatorname{Reset}(\hat{H}[t])=\left\{\begin{array}{rll}
\hat{V}_{\text {reset }}, & \text { if } \hat{H}[t] \geq \hat{V}_{\text {th}}^{\tau},;\\
\left\lfloor\frac{\theta-1}{\theta}\hat{H}[t]\right\rceil, & \text { if } \hat{V}_{\text {reset }} \leq \hat{H}[t]<\hat{V}_{\text {th }}^{\tau};\\
\hat{V}_{\text {reset }}, & \text { if } \hat{H}[t] \leq \hat{V}_{\text {reset }} .
\end{array}\right.
&
\end{aligned}\right.
\label{DiscretEquation}
\end{equation}

Here, the hat symbol represents the discretized values, and $\hat{V}_{\text {th}}^{\tau} = \theta \cdot \tau$. Moreover $\hat{I}[t]=\sum \hat{\omega}_{j} S_j[t].$
\end{theorem}

\begin{proof}
We multiply both sides of Equations of the LIF model(Eq.(\ref{LIF_equation})) by $\tau$ and the Sign function  has been substituted for the Fire function, which yields the following equations:

\begin{equation*}
\left\{
\begin{aligned}
\tau H[t] & = (\tau - 1)V[t-1] + I[t] ,\\
2S[t] & =\text { Sign }\left(\tau H[t]-V_{\text {th }}^{\tau}\right) + 1, \\
(\tau - 1)V[t] & =\operatorname{Reset}(\tau H[t])=\left\{\begin{array}{rll}
0, & \text { if } & \tau H[t] \geq V_{\text {th}}^{\tau}; \\
\frac{\tau-1}{\tau}(\tau H[t]), & \text { if } & 0 \leq \tau H[t] \leq V_{\text {th}}^{\tau}; \\
0, & \text { if } & \tau H[t] \leq 0 .
\end{array}\right.
\end{aligned}\right.
\end{equation*}
Then, we treat $\tau H[t]$ and $(\tau-1)V[t-1]$ as separate iterations objects. Therefore, we can rewrite $\tau H[t]$ as $H[t]$ without ambiguity, as well as $(\tau-1)V[t-1]$.

Finally, by multiplying the corresponding discretization factor $\theta$, we obtain Eq\ref{DiscretEquation}. Note that since the division operation has been moved to the Reset function, rounding is applied during discretization.

\end{proof}

In Eq.(\ref{DiscretEquation}), the Leaky Integrate-and-Fire (LIF) model degenerates into the Integrate-and-Fire (IF) model when $V_{\text{th}}^{\tau} = 1$ and $\tau = \infty$. To facilitate further discussions, we will refer to this set of equations as the LIF(IF) function.
\begin{equation}\label{LIF_IF}
\begin{aligned}
    2S[t] &= LIF(I[t]),\\
    2S[t] &= IF(I[t]).
\end{aligned}
\end{equation}

\subsection{Multi-Level Discretization}
In Eq \ref{LIF_IF}, LIF(IF) model twice spike signals. If left unaddressed, the next Spiking Activation Layer would receive twice the input. To tackle this issue, we propose a multi-level discretization method that can also resolve the division problem in average pooling.

First, we redefine the WeightSum as follows:
$$
I[t] = \sum \omega_{j}S_j[t] = \sum \frac{\omega_{j}}{2}{2S_j[t]}. \\
$$

Then, by reducing the scaling factor $\theta$ of the corresponding weights to $\theta/2$, we obtain that
$$
\begin{aligned}
\hat{I}[t] &\triangleq \sum \operatorname{Discret}(\frac{\omega_{j}}{2}, \theta) \cdot {2S_j} \\
&= \sum \operatorname{Discret}(\omega_{j}, \frac{\theta}{2}) \cdot {2S_j} \\
&\approx \theta \cdot I[t].\\
\end{aligned}
$$

This approach can be extended to the treatment of average pooling. The subsequent layer following the average pooling layer may consist of either a convolutional or a linear layer, which is subsequently fed into the Spiking Activation Layer. The computation involved in average pooling requires a division operation, which is not conducive to FHE. Consequently, we can apply a similar strategy by transferring this division operation to the subsequent linear layer. This process can be outlined as follows:
$$
I[t] = \sum_j \omega_{j}  \frac{\sum_k S_k[t]}{kn} = \sum_j\frac{\omega_{j}}{n}(\sum_k S_k[t]),\\
$$
where $\omega_{j}$ denotes the corresponding weight parameter and $n$ represents the divisor of the average pooling layer. Subsequently, by decreasing the scaling factor of the weights, $\theta$, to $\theta/n$, we obtain:

$$
\begin{aligned}
\hat{I}[t] &\triangleq \sum_j \operatorname{Discret}(\frac{\omega_{j}}{k}, \theta) \cdot (\sum_k S_k[t]) \\
&= \sum_j \operatorname{Discret}(\omega_{j}, \frac{\theta}{n}) \cdot (\sum_k S_k[t]) \\
&\approx \theta \cdot I[t].\\
\end{aligned}
$$

For each Spiking Activation Layer, the input is approximate $\theta$ times that of the original network, defined as scale-invariance. This property is crucial for FHE as it guarantees that each Spiking Activation Layer's message space is a multiple of $\theta$. By selecting suitable parameters, we can perform homomorphic evaluations on neural networks of any depth, independent of the network's depth.

% \newtheorem{proposition2}{Proposition}[section]
% \begin{proposition2}
%     123
% \end{proposition2}

% \begin{figure}
%     \centering
%     \includegraphics[width=0.8\textwidth]{figures/SGGN_Eigenvalue_Density.png}
%     \caption{Caption}
%     \label{fig:enter-label}
% \end{figure}

% The following Figure \ref{fig:enter-label} denotes

\section{Homomorphic Evaluation of DiCSNNs}
\label{sec:options}

In this chapter, we present FHE-DiCSNN, a network designed for performing forward propagation on ciphertexts. The chapter is divided into two parts. Firstly, we discuss the computation of convolutional layers, average pooling layers, and linear layers (fully connected layers) on ciphertext. While the WeightSum operation is inherently supported by FHE, it is crucial to carefully consider the maximum value and the growth of noise of ciphertexts during computation to avoid any potential errors. In the second part, we employ programmable bootstrapping techniques from \cite{chillotti2020tfhe} to homomorphically compute the Fire and Reset functions of LIF neurons, referred to as FHE-Fire and FHE-Reset functions respectively. The use of bootstrapping refreshes the ciphertext noise after each Spiking Activation Layer, eliminating the need for fixed constraints on the network depth. Thus, our framework offers flexibility in selecting network depths, facilitating the evaluation of neural networks with varying depths.

\subsection{Homomorphic Computation of WeightSum}

% Homomorphic Evaluation of Convolutional Layers, Linear Layers, and Average Pooling Layers
Weightsum performs the essential operation of multiplying the value vector of the lower layer by the weight vector and summing them up. The weights remain fixed during the prediction process. Essentially, WeightSum represents the dot product between the weight vector and the value vector of the input layer. In the ciphertext domain, this computation can be expressed as:
\begin{equation}
     \sum \hat{\omega}_{j} LWE(x_j) = LWE(\sum \hat{\omega}_{j} x_j) .\label{noise grows}
\end{equation}

Here, we omit the specific summation dimensions, which can be easily determined based on the convolutional layers, linear layers, and average pooling layers.

WeightSum is inherently supported by FHE. To ensure the correctness of the computation, representing as  $Dec(\sum \hat{\omega}_{j} LWE(x_j) ) = Dec(LWE(\sum \hat{\omega}_{j} x_j))$, two conditions must be satisfied: (1) $\sum_j \hat{\omega}_{j} x_j \in\left[-\frac{p}{2}, \frac{p}{2}\right)$; (2) The noise remains within the noise bound. The first condition can be easily fulfilled by selecting a large enough message space $\mathbb{Z}_p$.

Regarding the ciphertext noise, after the WeightSum operation, the noise grows to $\sum_j \left| \hat{\omega}_{j}\cdot \sigma\right|, $
assuming that $LWE(x_j)$ has an initial noise $\sigma$. This assumption is reasonable because $LWE(x_j)$ are generated by Spiking Activation Layers evaluated through bootstrapping. 

It is observed that the noise maximum is proportional to the discretization parameter $\theta$. One approach to control the noise growth is to decrease $\theta$, although this may result in reduced accuracy. Another strategy is to balance the security level by reducing the initial noise $\sigma$.

\subsection{Homomorphic Computation of LIF Neuron Model}
The Fire and Reset functions from the Eq\ref{DiscretEquation}, being non-polynomial functions, necessitate the utilization of programmable bootstrapping from Theorem \ref{Programtfhe} for computation. To address this, we propose the FHE-Fire and FHE-Reset functions, a framework specifically designed to implement the Fire and Reset functions on ciphertexts.

We define the program function $g$ as follows:
$$
g(m)=\left\{\begin{array}{rcc}
1, & \text { if } & m \in\left[0, \frac{p}{2}\right); \\
-1, & \text { if } & m \in\left[-\frac{p}{2}, 0\right).
\end{array}\right.
$$
Note that the condition $g\left(v+\frac{p}{2}\right)=-g(v)$ should be satisfied. Then, the Fire function can be represented as:
\begin{equation}
\begin{aligned}\label{FHE-Fire}
2 F H E-F i r e(L W E(m)) & =\text { bootstrap }(\operatorname{LWE}(m))+1 \\
& = \begin{cases}L W E(2), \text { if } & m \in\left[0, \frac{p}{2}\right); \\
L W E(0), \text { if } & m \in\left[-\frac{p}{2}, 0\right)\end{cases} \\
& =L W E(\operatorname{Sign}(m)+1) \\
& =L W E(2 \cdot \text { Spike }).
\end{aligned}
\end{equation}

Similar to the FHE-Fire function, the FHE-Reset function can be computed by defining the program function $g$ for bootstrapping as follows:
$$
g(m) \triangleq 
\begin{cases}
0, & \text { if } m \in\left[\hat{V}_{\text {th }}, \frac{p}{2}\right); \\
\left\lfloor\frac{\theta-1}{\theta}m\right\rceil, & \text { if } m \in\left[0, \hat{V}_{\text {th }}\right); \\ 
0, & \text { if } m \in\left[\hat{V}_{\text {th }}-\frac{p}{2}, 0\right); \\ 
\frac{p}{2}-\left\lfloor\frac{\theta-1}{\theta}m\right\rceil, & \text { if } m \in\left[-\frac{p}{2}, \hat{V}_{\text {th }}-\frac{p}{2}\right),\end{cases}
$$
where $g(x)=-g\left(x+\frac{p}{2}\right)$ must be satisfied too. Then, the FHE-Reset function can be computed as follows:
\begin{equation}\label{FHE-Reset}
\begin{aligned}
& \text { FHE-Reset }(L W E(m)) \triangleq \text { bootstrap }(L W E(m)) \\
& =\begin{cases}
L W E(0),& m \in\left[\hat{V}_{\text {th }}, \frac{p}{2}\right); \\
L W E(\left\lfloor\frac{\theta-1}{\theta}m\right\rceil),& m \in\left[0, \hat{V}_{\text {th }}\right); \\
L W E(0), &m \in\left[\hat{V}_{\text {th }}-\frac{p}{2}, 0\right); \\
L W E\left(\frac{p}{2}-\left\lfloor\frac{\theta-1}{\theta}m\right\rceil\right),& m \in\left[-\frac{p}{2}, \hat{V}_{\text {th }}-\frac{p}{2}\right).
\end{cases}\\
\end{aligned}
\end{equation}
Please note that if $m$ falls into the interval $\left[-\frac{p}{2}, \hat{V}_{\text {threshold }}-\frac{p}{2}\right)$, the FHE-Reset function will produce incorrect computation results. Therefore, we need to ensure that the value of $\hat{H}[t]$ does not fall into this interval. The following theorem demonstrates that this condition is easily satisfied.
\vspace{-0.1cm}
\begin{theorem}
\label{MAXMIN}    
If $\operatorname{M} \triangleq \hat{V}_{\text {th}}+\max _t(|\hat{I}[t]|)$ and $\operatorname{M} \leq \frac{p}{2}$, then $\hat{H}[t] \in\left[\hat{V}_{\text {threshold }}-\frac{p}{2}, \frac{p}{2}\right)$.
\end{theorem}

\begin{proof}
\begin{align}
\max (\hat{H}[t]) & =\max (\hat{V}[t]+\hat{I}[t]) \notag\\
& \leq \frac{\tau-1}{\tau} \hat{V}_{t h}+\max _t(|\hat{I}[t]|) \notag\\
& <\operatorname{M} \notag\\
& \leq \frac{p}{2} \label{maxH}\\
\min (\hat{H}[t]) & =\min (\hat{V}[t]+\hat{I}[t]) \notag\\
& \geq 0-\max _t(|\hat{I}[t]|) \notag\\
& \geq-\frac{p}{2}+\hat{V}_{t h} \notag\\
& \geq-\frac{p}{2}\label{minH}
\end{align}
\end{proof}

The above theorem states that as long as $\operatorname{M} \leq \frac{p}{2}$, the maximum and minimum values of $\hat{H}[t]$ will fall within the interval $\left[\hat{V}_{\text {threshold }}-\frac{p}{2}, \frac{p}{2}\right)$.
It can be readily demonstrated that the maximum value arising in the computation process of CSNN is guaranteed to occur in the variable $\hat{H}[t]$. This finding not only confirms the validity of the FHE-Reset function but also allows for an estimation of the maximum value within the message space. It also provides a convenient criterion for selecting the parameter $p$ for the message space.

Furthermore, the FHE-Fire and FHE-Reset functions not only compute Fire and Reset functions on ciphertexts but also refresh the ciphertext noise. This property is crucial as it ensures resulting ciphertexts have minimal initial noise. By keeping accumulated noise after linear layer operations below a predetermined upper bound, subsequent layers in CSNNs share the same initial noise, enabling accurate computations. In essence, our framework allows network expansion to arbitrary depths without noise concerns.

\section{Experiments}
In this chapter, we empirically demonstrate the excellent performance of FHE-DiCSNN in terms of accuracy and time efficiency. Firstly, we analyze the outstanding time efficiency of FHE-DiCSNN. Secondly, through theoretical analysis, we determine that the maximum value within the message space and the maximum noise growth are directly proportional to the discretization factor $\theta$. We design experiments to determine the corresponding proportionality coefficients, allowing us to select appropriate FHE parameters based on the value of $\theta$. Finally, we experimentally evaluate the actual accuracy and time efficiency of FHE-DiCSNN under different combinations of decay factor $\tau$ and discretization factor $\theta$.

\subsection{Time Consumption}
The structure of CSNN has been extensively discussed in Section 3. The convolutional layer plays a crucial role in extracting key image features, which, when combined with LIF neurons, enables pulse encoding specific to image features, replacing the stochastic Poisson encoding. If we replace the convolution process in CSNN, as shown in Figure \ref{CSNN_main}, with Poisson encoding \ref{Poisson-coding}, we obtain a fully connected SNN driven by Poisson encoding. However, Poisson encoding introduces randomness, and to obtain stable experimental results, a sufficiently large simulation time $T$ (which can be understood as the number of cycles for processing a single image) is required, significantly increasing the time consumption. In contrast, spiking encoding based on the convolutional layer can stably extract features, allowing the simulation time to be reduced to 2 cycles (to ensure that LIF neurons accumulate sufficient membrane potential to generate spikes).

The simulation time $T$ is a crucial factor that significantly affects time efficiency. It determines the number of cycles in the network and also increases the number of bootstrapping operations. Bootstrapping is the most time-consuming step in FHE, and in FHE-DiCSNN, it is reflected in the computation of LIF neurons. Therefore, the number of bootstrapping operations can be used as a simple estimate of time consumption. For each LIF neuron, two bootstrapping are required to compute FHE-Fire and FHE-Reset. On the other hand, Poisson encoding essentially involves one comparison and can be implemented using the Sign function, requiring one bootstrapping. The following table provides a simple estimation for the CSNN defined in Figure \ref{CSNN_main} and an equivalently dimensioned Poisson-encoded SNN:

\captionsetup[table]{position=bottom}

\begin{table}[H]
    \centering
    \resizebox{1\textwidth}{!}{
\begin{tabular}{ccc}
   \toprule
     & Poisson-encoded SNN & CSNN \\
    \midrule
   bootstrapping & $(784+2\times160+2\times10)\times T_1$ & $(1440\times 2+2\times 160+2\times 10)\times T_2$  \\
   Spiking Activation Layer & $5\times T_1$ & $6\times T_2$  \\
   \bottomrule
\end{tabular}}
\caption{$T_1$ and $T_2$ represent the simulation time required for Poisson-encoded SNN and CSNN, respectively, to achieve their respective peak accuracy performances. Typically, $T_1$ falls within the range of [20-100], while $T_2$ falls within the range of [2-4].}
\label{Poisson}
\end{table}

If we do not consider parallel computing, the number of bootstrapping can be used as a simple estimate of the network's time consumption. In this case, both Poisson-encoded SNN and CSNN would have a time consumption in the order of thousands. However, since the bootstrapping of Spiking Activation Layers and Poisson encoding can be performed in parallel, the time consumption will be proportional to the number of corresponding layers. In the case of parallel computing, CSNN exhibits a time efficiency that is 10 times higher than that of Poisson-encoded SNN.

\subsection{Parameters Selection}
In this part, we discuss the selection of FHE parameters. We begin with the message space $Z_p$. In the encryption scheme, $p$ acts as the modulus, ensuring all operations occur within the finite field $\mathbb{Z}_p$. It is crucial to monitor numerical growth and prevent subtraction operations from exceeding $p$ to avoid unexpected outcomes.
Theorem \ref{MAXMIN} provides an easy criterion to find the maximum value. As long as it is satisfied that
\begin{equation}
    \hat{V}_{\text{th}} + \max\limits_{t}(|\hat{I}[t]|) \leq \frac{p}{2},
\label{select_p}
\end{equation}
the value of the intermediate variable will not exceed the message space $\mathbb{Z}_p$. The formula
$$
\hat{V}_{\text{th}} + \max\limits_{t}(|\hat{I}[t]|) \approx \theta \cdot (V_{th}+\max_t (|I[t]|)) = \theta \cdot (V_{th} + \sum\omega_{j}S_j[t]),
$$
indicates that $\hat{I}$ is proportional to discretization parameter $\theta$.

We estimated the true maximum value of $V_{th} + \sum\omega_{j}S_j[t]$ on the training set, and the findings are summarized in Table \ref{SIZE_P}.

\begin{table}[H]
    \centering
    \resizebox{1\textwidth}{!}{
\begin{tabular}{cccc}
   \toprule
    $V_{th}+\max_t (|I[t]|)$ & Spiking Activation Layer1 & Spiking Activation Layer2 & Spiking Activation Layer3 \\
    \midrule
   $\tau=2$ & 29.03 & 9.25  & 4.93 \\
      $\tau=3$ & 36.13 & 10.85  & 6.44\\  
   $\tau=4$ & 2.8 & 0.96  & 0.17\\ 
      $\tau=\infty$(IF) & 23.00 & 11.64  & 6.39\\
   \bottomrule
    \end{tabular}}
    
    \caption{Under the given conditions of decay parameter $\tau = 2, 3, 4$, and $\tau=\infty (IF)$, we recorded the maximum values of the inputs to each Spiking Activation Layer during the forward propagation of the CSNNs network on the training set. After scaling the aforementioned values by $\theta$, we can estimate the maximum values that may arise in the FHE-DiCSNN. However, it is important to note that this estimation based on the training set parameters may lead to certain samples from the test set causing intermediate variables to exceed the predefined limits. Fortunately, the probability of such an occurrence is very low.}
\label{SIZE_P}
\end{table}

A technique was proposed to save computational cost 
by dynamically adjusting the size of the message space in the paper DiNN \cite{bourse2018fast}. This technique is also applicable to our work, so that a smaller plaintext space can be selected to reduce the growth rate of noise.

Accurately locating the noise growth is another problem we need to solve. The noise of the ciphertext only increases during the calculation of WeightSum. For a single WeightSum operation, since its inputs are ciphertexts with initial noise $\sigma$, the noise of the ciphertext will increase to
\begin{equation}
\sigma \sum_j \left| \hat{w}_{j}\right| \approx \theta \cdot \sigma \sum_j \left| w_{j}\right|.
\label{noise_estimate_computing}
\end{equation}

The above equation demonstrates that the maximum value of noise growth can be obtained by calculating $\max \sum_j |w_{j}|$, which can be determined at the time of setting because the weights are known. The experimental results are presented in the following table\ref{NOISEMAX}:

\begin{table}[H]
\caption{Based on the noise estimation formula mentioned above, the quantity $\max\sum_j \left|\omega_{j}\right|$ can be employed to estimate the growth of noise in DiCSNNs for various $\theta$.
}
\label{NOISEMAX}
    \centering
\begin{tabular}{ccccc}
   \toprule
   
    & $\tau=2$ & $\tau=3$ & $\tau=4$ & $\tau=\infty$(IF) \\
    \midrule
   $\max\sum_j | \omega_j|$ & 17.42 & 18.87 & 10.24 &  11.89 \\
   \bottomrule
\end{tabular}
\label{weight_sum_estimate}
\end{table}

%%%%%%加入一个介绍句
%From the From the two experiments: parameters $\tau$ and $\sigma$ selection
From the above discussion, it is evident that both the size of the message space and the upper bound of the noise exhibits a direct proportionality to $\theta$. The experimental results presented provide the corresponding scaling factors, enabling us to estimate the upper bounds associated with different $\theta$ values. With this information, suitable FHE parameters can be chosen or standard parameter sets such as STD128\cite{micciancio2021bootstrapping} can be utilized.

\subsection{Experimental Results}
\begin{figure}
    \centering
    \includegraphics[width=0.55\textwidth]{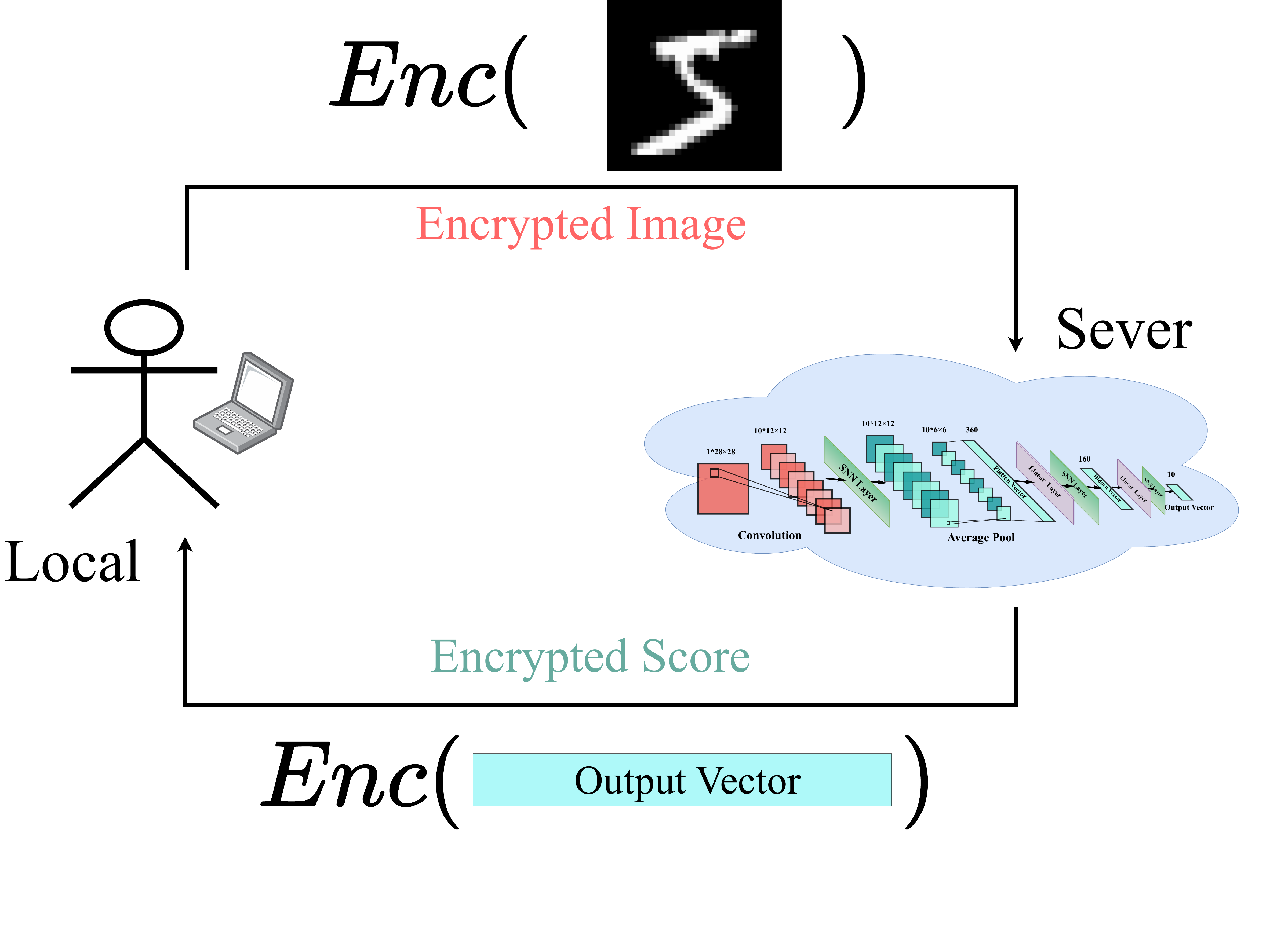}
    \caption{This figure showcases the practical application scenario of FHE-DiCSNN. The encrypted image, along with its corresponding evaluation key, is uploaded by the local user to the cloud server. Equipped with powerful computational capabilities, the server conducts network calculations. Subsequently, the classification scores are returned to the local user, who decrypts them to obtain the classification results.}
    \label{user_sever}
\end{figure}

Following the depicted process shown in Fig\ref{user_sever}, we conducted the experimental procedure during the noon time period using an Intel Core i7-7700HQ CPU @ 2.80 GHz. The procedure can be outlined as follows:

1. The grayscale handwritten digit image is encrypted into LWE ciphertext.

2. The ciphertext undergoes multiplication with discretized weights and is forwarded to the Spiking Activation Layer.

3. Within the Spiking Activation Layer, the LIF neuron model executes the FHE-Fire and FHE-Reset procedures on the ciphertext. Acceleration of bootstrapping operations is achieved through FFT technology and parallel computing.

4. Steps 1-3 are repeated $T$ times, and the resulting outputs are accumulated as classification scores.

5. Decryption is performed, and the highest score is selected as the classification result.
We selected different combinations of the decay parameter $\tau$ and scaling factor $\theta$, and the experimental results are presented in the Table \ref{tau}.

We have selected combinations of different decay parameters $\tau$ and discretization scaling factors $\theta$, and the experimental results are displayed in the following table:
 
\begin{table}[H]
\caption{During the evaluation of the FHE-DiCSNN network on the encrypted test set, we performed tests using different values of $\theta$. The last row showcases the performance of the original CSNNs on the plaintext test set, while the converged CSNNs was trained using the Spikingjelly\cite{SpikingJelly} framework.}
\label{EXRESULT}
    \centering
\begin{tabular}{ccccccc}
   \toprule
   
    & $\tau=2$ & $\tau=3$ & $\tau=4$ & $\tau=\infty$(IF) & Time/per image\\
    \midrule
   $\theta=10$ & 87.81\% & 75.62\%  &  8.49\% &  97.10\% & \multirow{3}{*}{0.75s}   \\
   $\theta=20$ & 92.67\%  & 76.98\% &  9.13\% & 97.67\% & \\
   $\theta=40$ & 94.77\%  & 79.35\%  &  9.80\%  & 97.94\% &  \\
   \hline
   CSNNs   &  95.53\%  &  89.94\% & 9.80\%  &  98.47\% &  \\
   \bottomrule
\end{tabular}
\label{tau}
\end{table}

The experimental findings highlight the significant negative impact of the decay factor $\tau$ on accuracy. Specifically, when $\tau=4$, the network becomes inactive. Analysis of the network's intermediate variables revealed a lack of spike generation by the neurons, and the weights in the second and third layers almost completely decay to zero. Thus, in CSNNs, ensuring the excitation of spikes is crucial. The size of the threshold voltage $\hat{V}_{th}^{\tau}$ directly influences spike generation, with larger $\hat{V}_{th}^{\tau}$ values making it more challenging to trigger spikes. Conversely, the IF model with the smallest $\hat{V}_{th}^{\tau}$ exhibits the highest accuracy.

On the other hand, the impact of $\theta$ on accuracy is positive, as larger $\theta$ values result in higher precision within the network. It is vital to emphasize that the choice of $\theta$ must be compatible with the size of the message space. Otherwise, an excessively large $\theta$ can cause the maximum value to exceed the range, leading to a detrimental effect on accuracy.

When selecting a smaller upper bound for noise, differences in spike generation frequency were observed between FHE-DiCSNN and DiCSNNs during network computations. This implies that certain ciphertexts may experience noise overflow, leading to incorrect classification results. However, this has a negligible impact on the final classification outcome. It occurs only at the edges of the threshold, where slight noise overflow happens with very low probability, resulting in occasional anomalous spike transitions. This intriguing experimental observation indicates that FHE-DiCSNN exhibits a certain level of noise tolerance.

\section{Conclusion}
This paper introduces the FHE-DiCSNN framework, which is built upon the efficient TFHE scheme and incorporates convolutional operations from CNN. The framework leverages the discrete nature of SNNs to achieve exceptional prediction accuracy and time efficiency in the ciphertext domain. The homomorphic computation of LIF neurons can be extended to other SNNs models, offering a novel solution for privacy protection in third-generation neural networks. Furthermore, by replacing Poisson encoding with convolutional methods, it improves accuracy and mitigates the issue of excessive simulation time caused by randomness. Parallelizing the bootstrapping computation through engineering techniques significantly enhances computational efficiency. Additionally, we provide upper bounds on the maximum value of homomorphic encryption and the growth of noise, supported by experimental results and theoretical analysis, which guide the selection of suitable homomorphic encryption parameters and validate the advantages of the FHE-DiCSNN framework.

There are also promising avenues for future research: 1. Exploring homomorphic computation of non-linear spiking neuron models, such as QIF and EIF. 2. Investigating alternative encoding methods to completely alleviate simulation time concerns for SNNs. 3. Exploring intriguing extensions, such as combining SNNs with RNNs or reinforcement learning and homomorphically evaluating these AI algorithms.

% \section*{Thanks}
% We would like to thank xxx.

% \appendix
% \section{The first part appendix}
% \subsection{ The first level of first part appendix}
% \section{The second part appendix}

\bibliographystyle{unsrt}
\bibliography{ref}

\end{document}